%% file: MainText.tex
\documentclass[%
 reprint,
 superscriptaddress,
 amsmath,amssymb,
 aps,
 prb,
 twocolumn
]{revtex4-2}

\newcommand{\tabincell}[2]{\begin{tabular}{@{}#1@{}}#2\end{tabular}}
\usepackage{graphicx}
\usepackage{dcolumn}
\usepackage{amsmath}
\usepackage{bm}
\usepackage{colortbl}
\usepackage{float}
\usepackage[colorlinks=true, urlcolor=blue, linkcolor=blue, citecolor=blue]{hyperref}
\usepackage[table,dvipsnames]{xcolor}
\usepackage{makecell}
\usepackage[mathscr]{euscript}
\usepackage{multirow}
\usepackage{braket}
\usepackage[pagewise]{lineno}
\usepackage{enumitem}
\usepackage{supertabular}
\usepackage{amssymb}
\usepackage{makecell}
\usepackage{tikz}
\usepackage{bbding}

\allowdisplaybreaks[4]
\begin{document}

\title{Anomalous Hall effect from nonlinear magnetoelectric coupling}

\author{Longju Yu}
\affiliation{Key Laboratory of Material Simulation Methods and Software of Ministry of Education, College of Physics, Jilin University, Changchun 130012, China}

\author{Hong Jian Zhao}
\affiliation{Key Laboratory of Material Simulation Methods and Software of Ministry of Education, College of Physics, Jilin University, Changchun 130012, China}
\affiliation{Key Laboratory of Physics and Technology for Advanced Batteries (Ministry of Education), College of Physics, Jilin University, Changchun 130012, China}
\affiliation{International Center of Future Science, Jilin University, Changchun 130012, China}

\author{Yurong Yang}
\affiliation{National Laboratory of Solid State Microstructures, Nanjing University, Nanjing 210093, China}
\affiliation{Jiangsu Key Laboratory of Artificial Functional Materials, Department of Materials Science and Engineering, Nanjing University, Nanjing 210093, China}

\author{Laurent Bellaiche}
 \affiliation{Smart Ferroic Materials Center, Physics Department and Institute for Nanoscience and Engineering, University of Arkansas, Fayetteville, Arkansas 72701, USA}
\affiliation{Department of Materials Science and Engineering, Tel Aviv University, Ramat Aviv, Tel Aviv 6997801, Israel}

\author{Yanming Ma}
\affiliation{Key Laboratory of Material Simulation Methods and Software of Ministry of Education, College of Physics, Jilin University, Changchun 130012, China}
\affiliation{International Center of Future Science, Jilin University, Changchun 130012, China}
\affiliation{State Key Laboratory of High Pressure and Superhard Materials, College of Physics, Jilin University, Changchun 130012, China}

\begin{abstract}
The anomalous Hall effect (AHE) is a topology-related transport phenomenon being of potential interest in spintronics, because this effect enables the efficient probe of magnetic orders (i.e., data readout in memory devices). It is well known that AHE spontaneously occurs in ferromagnets or antiferromagnets with magnetization. While recent studies reveal electric-field induced AHE (via linear magnetoelectric coupling), an AHE originating from {\it nonlinear} magnetoelectric coupling remains largely unexplored. Here, by symmetry analysis, we establish the phenomenological theory regarding the spontaneous and electric-field driven AHE in magnets. We show that a large variety of magnetic point groups host an AHE that is driven by uni-axial, bi-axial, or tri-axial electric field and that comes from nonlinear magnetoelectric coupling. Such electric-field driven anomalous Hall conductivities are reversible by reversing the magnetic orders. Furthermore, our first-principles calculations suggest Cr$_2$O$_3$ and CoF$_2$ as candidates hosting the aforementioned AHE. Our work emphasizes the important role of nonlinear magnetoelectric coupling in creating exotic transport phenomena, and offers alternative avenues for the probe of magnetic orders.
\end{abstract}

\maketitle

\section{Introduction}

The anomalous Hall effect (AHE) is a fundamental transport phenomenon where an applied voltage induces an transverse Hall current response~\cite{xiao2010berry,PhysRevLett.88.207208,onoda2002topological,bellaiche2013coupling,nagaosa2010anomalous,karplus1954hall,smit1955spontaneous,smit1958spontaneous,PhysRevB.2.4559,betancourt2023spontaneous,nayak2016large,nakatsuji2015large,vsmejkal2020crystal,li2023field,vsmejkal2022anomalous,reichlova2024observation,wang2023emergent,ikhlas2022piezomagnetic}. Nowadays, the AHE not only becomes an important platform for exploring topology-related transport properties~\cite{xiao2010berry,PhysRevLett.88.207208,onoda2002topological,bellaiche2013coupling,nagaosa2010anomalous}, but also offers an efficient way to probe magnetic order --- being of potential value in spintronics~\cite{han2024electrical,feng2022anomalous,gonzalez2023spontaneous,ikhlas2022piezomagnetic}. The AHE was found to spontaneously occur in ferromagnets~\cite{xiao2010berry,PhysRevLett.88.207208,bellaiche2013coupling,onoda2002topological,nagaosa2010anomalous,karplus1954hall,smit1955spontaneous,smit1958spontaneous,PhysRevB.2.4559} or antiferromagnets with specific magnetic configurations~\cite{betancourt2023spontaneous,nayak2016large,nakatsuji2015large,vsmejkal2020crystal,li2023field,vsmejkal2022anomalous,reichlova2024observation,wang2023emergent,ikhlas2022piezomagnetic}. As a matter of fact, antiferromagnetic materials with symmetry-allowed weak or tiny magnetization likewise enable AHE~\cite{grimmer1993general,vsmejkal2020crystal,Seemann2015SymmetryImposed}. By this, it is anticipated that external fields capable of inducing magnetization should, in principle, induce AHE as well. In this regard, magnetoelectric coupling (linear or nonlinear effect)~\cite{fiebig2005revival,schmid1973magnetoelectric,Hans_ME} are interesting avenues for electric-field driven AHE. Indeed, recent studies demonstrate that gate voltage can induce AHE via linear magnetoelectric coupling (e.g., the layer Hall effect and electric Hall effect~\cite{PhysRevB.106.245425,chen2024layer,gao2021layer,cui2024electrichalleffectquantum,PhysRevLett.133.096803}); Yet, the electric-field induced AHE arising from nonlinear magnetoelectric coupling remains largely unexplored.

Here, we use group-theoretical approach to establish the phenomenological theory for spontaneous and electric-field driven AHE in crystalline materials. Our symmetry analysis demonstrates that both type-I and type-III magnetic point groups (MPGs)~\footnote{MPGs are categorized into three types, namely (i) type-I MPGs involving only spatial operations and lacking time-reversal operation; (ii) type-II MPGs incorporating time-reversal as an independent operation; and (iii) type-III MPGs containing time-reversal operation only in combination with spatial operations. More details are provided in Ref.~\cite{dresselhaus2007group}.} exhibit spontaneous AHE or electric-field induced AHE from linear or nonlinear magnetoelectric coupling. We highlight that nonlinear magnetoelectric AHE is accommodated by a wide spectrum of MPGs, where the anomalous Hall conductivity component is (i) achieved by applying uni-axial, bi-axial, or tri-axial electric field and (ii) reversible by reversing the magnetic orders. Following our theory and confirmed by first-principles calculations, we further suggest that Cr$_2$O$_3$ and CoF$_2$ are representative materials towards the electric-field induced AHE via nonlinear magnetoelectric coupling.  \\

\section{Methods}

The Vienna Ab-initio Simulation Package (VASP)~\cite{PhysRevB.54.11169,PhysRevB.59.1758} was employed for first-principles calculations. Focusing on CoF$_2$ and Cr$_2$O$_3$, our calculations were based on the projector augmented wave (PAW) potential~\cite{PhysRevB.50.17953} and LDA functional~\cite{PhysRevLett.45.566} plus Hubbard $U$ corrections (Dudarev approach)~\cite{PhysRevB.57.1505}. During our calculations, the following electrons were considered: $(3s3p3d4s)$ for Co, $(3s3p3d4s)$ for Cr, $(2s2p)$ for F, and $(2s2p)$ for O. The effective Hubbard $U$ of 3.0 eV was used for the $3d$ electrons of Co and Cr ions. The kinetic cutoff energy of 550 eV, and $k$-point grids of $10\times10\times14$ for CoF$_2$ and $8\times8\times8$ for Cr$_2$O$_3$~\footnote{The magnetic space group of Cr$_2$O$_3$ is $R\bar{3}^\prime c^\prime$. In our calculations, the rhombohedral setting (10-atom cell) is adopted for Cr$_2$O$_3$.} were used during the structural relaxations and self-consistent calculations. The structural relaxations were carried out in the framework of collinear magnetism without involving spin-orbit coupling, with the force convergence criterion of 5 meV/\AA. The approaches mentioned in Refs.~\cite{chen2019electric,chen2016giant,fu2003first} allow us to determine the crystal structures of the crystalline materials under finite electric fields. In self-consistent-based transport calculations, noncollinear magnetism and the spin-orbit coupling were considered.

Starting from the optimized crystal structures of CoF$_2$ and Cr$_2$O$_3$ under zero and finite electric fields, we calculated the anomalous Hall conductivities by the maximally localized Wannier functions (MLWFs) approach --- implemented in the Wannier90 code~\cite{marzari1997maximally,marzari2012maximally}. The MLWFs were generated by considering (i) Co's $d$ orbitals and F's $p$ orbitals for CoF$_2$, and (ii) Cr's $d$ orbitals and O's $p$ orbitals for Cr$_2$O$_3$. The anomalous Hall conductivities were then computed with denser $k$-point grids of $150\times150\times210$ for CoF$_2$ and $100\times100\times100$ for Cr$_2$O$_3$.

Besides VASP and Wannier90, several other codes or databases were also employed, including Mathematica~\cite{mathematica}, Bilbao Crystallographic Server~\cite{aroyo2011crystallography,aroyo2006bilbao2,aroyo2006bilbao} (including Mpoint and Magndata~\cite{perez2015symmetry,gallego2016magndata,gallego2016magndata2}), Findsym~\cite{stokes2005findsym}, Seekpath~\cite{hinuma2017band,togo2018texttt}, Vesta~\cite{Momma2011VESTA3F}, Vaspkit~\cite{VASPKITCPC} and Matplotlib~\cite{Hunter:2007}.

\section{Results and discussion} 

\subsection{Anomalous Hall effect and Hall vector}

The AHE is characterized by $J_{\alpha} = \sum_{\beta}\sigma_{\alpha\beta}\mathcal{E}_{\beta}$, with $\sigma_{\alpha\beta}$ being the transverse conductivity (i.e., $\alpha\neq\beta$), $\mathcal{E}_{\beta}$ the electric field along $\beta$ direction, and $J_\alpha$ the current density along $\alpha$ direction. Generally, the anomalous Hall conductivity $\sigma_{\alpha\beta}$ stems from intrinsic~\cite{betancourt2023spontaneous,nayak2016large,nakatsuji2015large,li2023field,vsmejkal2020crystal} and extrinsic mechanisms~\cite{smit1955spontaneous,smit1958spontaneous,PhysRevB.2.4559}. This work focuses on the intrinsic mechanism. According to Refs.~\cite{nagaosa2010anomalous,vsmejkal2022anomalous}, the $\sigma_{\alpha\beta}$ conductivity is expressed as
\begin{equation}\label{Hall_eq}
\sigma_{\alpha\beta}=-\sum_{n\gamma}\epsilon_{\alpha \beta \gamma}\frac{e^2}{\hbar}\int \frac{d\mathbf{k}}{(2\pi)^3}f(\epsilon_{n\mathbf{k}})\Omega_{\gamma,n\mathbf{k}},
\end{equation}
where $\epsilon_{\alpha \beta \gamma}$ is the Levi-Civita symbol ($\alpha,\beta,\gamma=x,y,z$), $e$ the charge of the electron, $\hbar$ the reduced Planck constant, $\epsilon_{n\mathbf{k}}$ energy eigenvalue, $f(\epsilon_{n\mathbf{k}})$ the Fermi-Dirac distribution function, and $\Omega_{\gamma,n\mathbf{k}}$ the Berry curvature ($n$ being the band index, $\mathbf{k}$ being the wave vector). In Eq.~(\ref{Hall_eq}), the Levi-Civita symbol $\epsilon_{\alpha \beta \gamma}$ implies the anti-symmetric nature of the anomalous Hall conductivity, that is, $\sigma_{\alpha\beta}=-\sigma_{\beta\alpha}$. This enables the definition of a Hall vector $\boldsymbol{\mathcal{M}} \equiv (\mathcal{M}_x, \mathcal{M}_y, \mathcal{M}_z) \equiv (\sigma_{zy}, \sigma_{xz}, \sigma_{yx})$, a current density vector $\mathbf{J}\equiv(J_x,J_y,J_z)$ and an electric field vector $\boldsymbol{\mathcal{E}}\equiv(\mathcal{E}_x,\mathcal{E}_y,\mathcal{E}_z)$ so that the AHE is described by $\mathbf{J} =\boldsymbol{\mathcal{M}} \times \boldsymbol{\mathcal{E}}$ ~\cite{vsmejkal2022anomalous,vsmejkal2020crystal}. From symmetry point of view, the Hall vector $\boldsymbol{\mathcal{M}}$ behaves like the magnetization vector $\mathbf{M}$ [see Refs.~\cite{grimmer1993general,vsmejkal2020crystal,Seemann2015SymmetryImposed} and Section I of Supplementary Material (SM)~\footnote{See Supplemental Material for  the discussions on the anomalous Hall conductivity and Hall vector, symmetry analysis, and some numerical results, which includes Refs.~\cite{vsmejkal2022anomalous,xiao2010berry,grimmer1993general,Seemann2015SymmetryImposed,mtensor,mpoint,shtrikman1965remarks,Campbell:ib5106,nagaosa2010anomalous,vsmejkal2020crystal,fiebig2005revival,Hans_ME} } for details]. This establishes the correspondence between $(\sigma_{zy}, \sigma_{xz}, \sigma_{yx})$ and $(M_x, M_y, M_z)$.

\begin{table*}[htb]
\centering
\renewcommand{\arraystretch}{1.5}
\setlength{\tabcolsep}{0.5mm}{
\caption{\label{tab:SymmetryBreaking}The anomalous Hall conductivity in type-I and type-III MPGs. We use different symbols to label the anomalous Hall conductivity components that are spontaneously arisen or induced by electric fields. Specifically, the spontaneous components are indicated by the ``$\checkmark$'' symbol; The components driven by uni-axial $E_x$, $E_y$, and $E_z$ electric fields are indicated by ``$1$'', ``$2$'', and ``$3$'', respectively; The components driven by bi-axial $[E_x, E_y]$, $[E_x, E_z]$, and $[E_y, E_z]$ electric fields are labeled as ``$4$'', ``$5$'', and ``$6$'', respectively; The components induced by a tri-axial $[E_x, E_y, E_z]$ electric field is labeled as `$7$''. In cases where a conductivity component responds to multiple mechanisms, all corresponding numbers are listed within the relevant entries. As an example, the $\bar{6}m^\prime2^\prime$ MPG allows a spontaneous $\sigma_{xy}$ component, a $\sigma_{zy}$ component induced by bi-axial $[E_x,E_z]$ field, and a $\sigma_{xz}$ component induced by bi-axial $[E_x,E_z]$ or $[E_y,E_z]$ field.}

 \begin{tabular}{b{1.4cm}<{\centering}|b{0.57cm}<{\centering}b{0.57cm}<{\centering}b{0.57cm}<{\centering}|b{1.4cm}<{\centering}|b{0.57cm}<{\centering}b{0.57cm}<{\centering}b{0.57cm}<{\centering}|b{1.4cm}<{\centering}|b{0.57cm}<{\centering}b{0.57cm}<{\centering}b{0.57cm}<{\centering}|b{1.4cm}<{\centering}|b{0.57cm}<{\centering}b{0.57cm}<{\centering}b{0.57cm}<{\centering}|b{1.4cm}<{\centering}|b{0.57cm}<{\centering}b{0.57cm}<{\centering}b{0.57cm}<{\centering}}
    \toprule
MPGs&$\sigma_{zy}$&$\sigma_{xz}$&$\sigma_{yx}$&MPGs&$\sigma_{zy}$&$\sigma_{xz}$&$\sigma_{yx}$&MPGs&$\sigma_{zy}$&$\sigma_{xz}$&$\sigma_{yx}$&MPGs&$\sigma_{zy}$&$\sigma_{xz}$&$\sigma_{yx}$&MPGs&$\sigma_{zy}$&$\sigma_{xz}$&$\sigma_{yx}$\\
\hline
$1.1$&\checkmark&\checkmark&\checkmark&
$\bar{1}.1$&\checkmark&\checkmark&\checkmark&
$\bar{1}^\prime$&$123$&$123$&$123$&
$2.1$&$12$&$12$&\checkmark&$2^\prime$&\checkmark&\checkmark&$12$\\
\hline
$m.1$&$3$&$3$&\checkmark&$m^\prime$&\checkmark&\checkmark&$3$&$2/m.1$&$56$&$56$&\checkmark&
$2^\prime/m$&$3$&$3$&$12$&
$2/m^\prime$&$12$&$12$&$3$\\
\hline
$2^\prime/m^\prime$&\checkmark&\checkmark&$56$&$222.1$&$1$&$2$&$3$&$2^\prime2^\prime 2$&$2$&$1$&\checkmark&
$mm2.1$&$2$&$1$&$4$&
$m^\prime m2^\prime$&$4$&\checkmark&$2$\\
\hline
$m^\prime m^\prime2$&$1$&$2$&\checkmark&
$mmm.1$&$6$&$5$&$4$&
$m^\prime mm$&$7$&$3$&$2$&
$m^\prime m^\prime m$&$5$&$6$&\checkmark&
$m^\prime m^\prime m^\prime$&$1$&$2$&$3$\\
\hline
$4.1$&$12$&$12$&\checkmark&
$4^\prime$&$12$&$12$&$12$&
$\bar{4}.1$&$12$&$12$&\checkmark&
$\bar{4}^\prime$&$12$&$12$&$123$&
$4/m.1$&$56$&$56$&\checkmark\\
\hline
$4^\prime/m$&$56$&$56$&$12$&
$4/m^\prime$&$12$&$12$&$3$&
$4^\prime/m^\prime$&$12$&$12$&$56$&
$422.1$&$1$&$2$&$3$&
$4^\prime 2 2^\prime$&$1$&$2$&$456$\\
\hline
$4 2^\prime 2^\prime$&$2$&$1$&\checkmark&
$4mm.1$&$2$&$1$&$4$&
$4^\prime m^\prime m$&$1$&$2$&$12$&
$4 m^\prime m^\prime$&$1$&$2$&\checkmark&
$\bar{4}2m.1$&$1$&$2$&$456$\\
\hline
$\bar{4}^\prime2^\prime m$&$2$&$1$&$12$&
$\bar{4}^\prime2m^\prime$&$1$&$2$&$3$&
$\bar{4}2^\prime m^\prime$&$2$&$1$&\checkmark&
$4/mmm.1$&$6$&$5$&$4$&
$4/m^\prime mm$&$2$&$1$&$7$\\
\hline
$4^\prime/mm^\prime m$&$5$&$6$&$12$&
$4^\prime/m^\prime m^\prime m$&$1$&$2$&$56$&
$4/mm^\prime m^\prime$&$5$&$6$&\checkmark&
$4/m^\prime m^\prime m^\prime$&$1$&$2$&$3$&
$3.1$&$12$&$12$&\checkmark\\
\hline
$\bar{3}.1$&$12$&$12$&\checkmark&
$\bar{3}^\prime$&$12$&$12$&$123$&
$32.1$&$12$&$2$&$23$&
$32^\prime$&$2$&$12$&\checkmark&
$3m.1$&$12$&$1$&$1$\\
\hline
$3m^\prime$&$1$&$12$&\checkmark&
$\bar{3}m.1$&$12$&$45$&$45$&
$\bar{3}^\prime m$&$2$&$1$&$1$&
$\bar{3}^\prime m^\prime$&$1$&$2$&$23$&
$\bar{3}m^\prime$&$45$&$12$&\checkmark\\
\hline
$6.1$&$12$&$12$&\checkmark&
$6^\prime$&$12$&$12$&$12$&
$\bar{6}.1$&$56$&$56$&\checkmark&
$\bar{6}^\prime$&$12$&$12$&$3$&
$6/m.1$&$56$&$56$&\checkmark\\
\hline
$6^\prime/m$&$56$&$56$&$12$&
$6/m^\prime$&$12$&$12$&$3$&
$6^\prime/m^\prime$&$12$&$12$&$56$&
$622.1$&$1$&$2$&$3$&
$6^\prime22^\prime$&$12$&$456$&$2$\\
\hline
$62^\prime 2^\prime$&$2$&$1$&\checkmark&
$6mm.1$&$2$&$1$&$4$&
$6^\prime m m^\prime$&$12$&$4$&$1$&
$6m^\prime m^\prime$&$1$&$2$&\checkmark&
$\bar{6}m2.1$&$56$&$5$&$1$\\
\hline
$\bar{6}^\prime m^\prime 2$&$1$&$12$&$3$&
$\bar{6}^\prime m2^\prime$&$12$&$1$&$5$&
$\bar{6}m^\prime2^\prime$&$5$&$56$&\checkmark&
$6/mmm.1$&$6$&$5$&$4$&
$6/m^\prime m m$&$2$&$1$&$7$\\
\hline
$6^\prime/mmm^\prime$&$56$&$7$&$1$&
$6^\prime/m^\prime mm^\prime$&$12$&$4$&$5$&
$6/mm^\prime m^\prime$&$5$&$6$&\checkmark&
$6/m^\prime m^\prime m^\prime$&$1$&$2$&$3$&
$23.1$&$1$&$2$&$3$\\
\hline
$432.1$&$1$&$2$&$3$&
$4^\prime32^\prime$&$456$&$456$&$456$&
$\bar{4}3m.1$&$456$&$456$&$456$&
$\bar{4}^\prime3m^\prime$&$1$&$2$&$3$&
$m\bar{3}.1$&$6$&$5$&$4$\\
\hline
$m^\prime \bar{3}^\prime$&$1$&$2$&$3$&
$m\bar{3}m.1$&$6$&$5$&$4$&
$m^\prime\bar{3}^\prime m$&$45$&$46$&$56$&
$m\bar{3}m^\prime$&$6$&$5$&$4$&
$m^\prime\bar{3}^\prime m^\prime$&$1$&$2$&$3$\\
 \hline
 \hline
 \end{tabular}}
\end{table*}

For a material with magnetic order parameter $L$, the $\sigma_{\alpha\beta}$ conductivity depends on the orientation of $L$. To show this, we rewrite $\sigma_{\alpha\beta}$ as $\sigma_{\alpha\beta}(L)$. We recall that time-reversal symmetry links $+L$ with $-L$ magnetic order parameter. The Onsager reciprocity relation further requires that 
$\sigma_{\alpha\beta}(+L)=\sigma_{\beta\alpha}(-L)$~\cite{Seemann2015SymmetryImposed,grimmer1993general} and this yields
\begin{equation}\label{onsager}
\sigma_{\alpha\beta}(+L)=-\sigma_{\alpha\beta}(-L),
\end{equation}
according to the anti-symmetric feature of $\sigma_{\alpha\beta}$. This indicates that the anomalous Hall conductivity can be reversed by reversing the magnetic order parameter.\\

\subsection{Anomalous Hall effect in magnets}

We move on to explore the AHE in crystalline materials. Because of the linkage between AHE and magnetization, we are interested in the magnetization in materials that is spontaneously existing or induced by external stimuli. According to Refs.~\cite{dresselhaus2007group,schmid1973magnetoelectric}, the existence or absence of magnetization in crystalline materials is determined by their magnetic point groups. For instance, the $m^\prime m2^\prime$ MPG contains 4 symmetry operations, namely, an identity operation ($\mathfrak{1}$), a mirror plane perpendicular to $x$ followed by a time-reversal ($\mathfrak{m}_x^\prime$), a mirror plane perpendicular to $y$ ($\mathfrak{m}_y$), and a twofold rotation along $z$ followed by a time-reversal ($\mathfrak{2}_z^\prime$). The $\mathfrak{1}$, $\mathfrak{m}_x^\prime$, $\mathfrak{m}_y$ and $\mathfrak{2}_z^\prime$ operations transform $(M_x,M_y,M_z)$ to $(M_x,M_y,M_z)$, $(-M_x,M_y,M_z)$, $(-M_x,M_y,-M_z)$, and $(M_x,M_y,-M_z)$, respectively. This means that $M_y$ is invariant under the 4 symmetry operations, and the $m^\prime m2^\prime$ MPG enables $M_y$ magnetization.

If an MPG is not compatible with $M_\alpha$ magnetization, there must be one or more symmetry operations in this MPG forbidding $M_\alpha$. An avenue towards $M_\alpha$ is to break such symmetry operations by external fields. Thanks to magnetoelectric coupling~\cite{fiebig2005revival,schmid1973magnetoelectric,Hans_ME}, applying external electric field $\mathbf{E}=(E_x,E_y,E_z)$ might induce magnetization. To demonstrate this, we again take the $m^\prime m2^\prime$ MPG as an example. In $m^\prime m2^\prime$ MPG, both $\mathfrak{m}_y$ and $\mathfrak{2}_z^\prime$ symmetry operations forbid $M_z$ magnetization, because $\mathfrak{m}_y: M_z \rightarrow -M_z$ and $\mathfrak{2}_z^\prime: M_z \rightarrow -M_z$. Applying a uni-axial electric field $E_y$ naturally breaks $\mathfrak{m}_y$ and $\mathfrak{2}_z^\prime$ symmetries (i.e., $\mathfrak{m}_y: E_y \rightarrow -E_y$, $\mathfrak{2}_z^\prime: E_y \rightarrow -E_y$), which generates an $M_z$ magnetization. By comparison, applying a uni-axial $E_x$, $E_y$, or $E_z$ electric field is insufficient to induce $M_x$ magnetization. As a matter of fact, $M_x$ is forbidden by $\mathfrak{m}_x^\prime$ and $\mathfrak{m}_y$, with $\mathfrak{m}_x^\prime: (E_x, E_y, E_z)\rightarrow(-E_x, E_y, E_z)$ and $\mathfrak{m}_y: (E_x, E_y, E_z)\rightarrow(E_x, -E_y, E_z)$. The $E_x$ electric field preserves $\mathfrak{m}_y$ symmetry, $E_y$ preserves $\mathfrak{m}_x^\prime$, and $E_z$ preserves both $\mathfrak{m}_y$ and $\mathfrak{m}_x^\prime$. In such sense, the uni-axial $E_\alpha$ ($\alpha$ being $x$, $y$, or $z$) electric field preserves either $\mathfrak{m}_x^\prime$ or $\mathfrak{m}_y$ symmetry, and this forbids $M_x$. Applying a bi-axial electric field $[E_x, E_y]$ simultaneously breaks $\mathfrak{m}_x^\prime$ and $\mathfrak{m}_y$ symmetries, yielding the $M_x$ magnetization.

By similar procedures, we perform symmetry analysis on the entire 122 MPGs, as detailed in Section II of SM. Among these MPGs, 32 type-II MPGs exhibit time-reversal symmetry $\hat{\mathcal{T}}$, which transforms $M_\alpha$ to $-M_\alpha$ ($\alpha=x,y,z$) and forbids the magnetization. Since electric field does not break $\hat{\mathcal{T}}$ symmetry, type-II MPGs do not host electric-field induced magnetization. We therefore omit the discussion on type-II MPGs. As for type-I and type-III MPGs, spontaneous and/or electric-field driven magnetization may be allowed by symmetry. Further, the occurrence of $M_x$, $M_y$, and $M_z$ implies $\sigma_{zy}$, $\sigma_{xz}$, and $\sigma_{yx}$ conductivities, respectively (see the previous section). In Table~\ref{tab:SymmetryBreaking}, we list the 90 type-I and type-III MPGs with respect to the spontaneous or electric-field driven anomalous Hall conductivity components.\\

\subsection{Phenomenological theory}

In this section, we establish the phenomenological theory for the spontaneous and electric-field driven AHE. We work with the magnetization component $M_\alpha$, and recall that $M_\alpha$ is rooted in the $B^{\alpha}_\mathrm{eff} M_\alpha$ coupling --- $B^{\alpha}_\mathrm{eff}$ being an effective magnetic field along $\alpha$ direction. The effective $B^{\alpha}_\mathrm{eff}$ field can spontaneously occur or be driven by uni-axial $E_\beta$, bi-axial $[E_\beta,E_\gamma]$ or tri-axial $[E_\beta,E_\gamma,E_\delta]$ electric field. Such 4 types of effective magnetic field $B^{\alpha}_\mathrm{eff}$ can be respectively written as
 \begin{equation}\label{h_alpha_expand}
 \begin{split}    &B^{\alpha,0}_\mathrm{eff}=\lambda_{\alpha},\\
 &B^{\alpha,\beta}_\mathrm{eff}=\sum_{l}\lambda_{\alpha\beta,l}E_{\beta}^l,\\
 &B^{\alpha,\beta\gamma}_\mathrm{eff}=\sum_{lm}\lambda_{\alpha\beta\gamma,lm}E_{\beta}^lE_{\gamma}^m,\\
 &B^{\alpha,\beta\gamma\delta}_\mathrm{eff}=\sum_{lmn}\lambda_{\alpha\beta\gamma\delta,lmn}E_{\beta}^lE_{\gamma}^mE_{\delta}^n,\\
 \end{split}
\end{equation}
with $\alpha$, $\beta$, $\gamma$, and $\delta$ labeling the Cartesian directions. The $\lambda_{\alpha}$ coefficient in Eq.~(\ref{h_alpha_expand}) implies the possible spontaneous $M_\alpha$ magnetization, while $\lambda_{\alpha\beta,l}$, $\lambda_{\alpha\beta\gamma,lm}$, and $\lambda_{\alpha\beta\gamma\delta,lmn}$ (with $l$, $m$, and $n$ being positive integers) correspond to linear or nonlinear magnetoelectric couplings.

Essentially, the analytical form of effective magnetic field is determined by symmetry. We derive the effective magnetic fields associated with the 90 type-I and type-III MPGs (see Appendix A for details) and summarize the results in Table~\ref{tab:coupling1} of the Appendix A. For several MPGs, the effective magnetic fields are not null even in the absence of external electric fields. This is exemplified by the $\bar{1}.1$ MPG with $B^{x,0}_\mathrm{eff}=\lambda_{x}$, $B^{y,0}_\mathrm{eff}=\lambda_{y}$, and $B^{z,0}_\mathrm{eff}=\lambda_{z}$ (see Table~\ref{tab:coupling1} of the Appendix A). Hence, $M_x$, $M_y$, and $M_z$ (i.e., $\sigma_{zy}$, $\sigma_{xz}$, $\sigma_{yx}$) components are enabled in $\bar{1}.1$ MPG. In other cases, the effective magnetic fields are driven by external electric field. As shown in Table~\ref{tab:coupling1} of the Appendix A, the effective magnetic fields associated with $\bar{6}m2.1$ MPG are given by
\begin{equation}\label{B6m11}
 \begin{split}    
&B^{x,xz}_\mathrm{eff}=\lambda_{xxz,21}E_x^2E_z,\\
&B^{x,yz}_\mathrm{eff}=\lambda_{xyz,11}E_yE_z,\\ 
&B^{y,xz}_\mathrm{eff}=\lambda_{yxz,11}E_xE_z,\\
&B^{z,x}_\mathrm{eff}=\lambda_{zx,3}E_x^3.
 \end{split}
\end{equation}
The $B^{x}_\mathrm{eff}$ field can be contributed by $B^{x,xz}_\mathrm{eff}$ and $B^{x,yz}_\mathrm{eff}$, where (i) $\lambda_{xxz,21}E_x^2E_z$ implies the third-order magnetoelectric coupling created by $[E_x, E_z]$ bi-axial electric field, and (ii) $\lambda_{xyz,11}E_yE_z$ implies the second-order magnetoelectric coupling due to $[E_y, E_z]$. Similarly, $B^{y}_\mathrm{eff}$ and $B^{z}_\mathrm{eff}$ can be contributed by $B^{y,xz}_\mathrm{eff}$ and $B^{z,x}_\mathrm{eff}$, which are induced by bi-axial $[E_x, E_z]$ and uni-axial $E_x$, respectively. In short, the $\bar{6}m2.1$ MPG enables $\sigma_{zy}$, $\sigma_{xz}$, and $\sigma_{yx}$, where $\sigma_{zy}$ responds to $E_x^2 E_z$ or $E_y E_z$, $\sigma_{xz}$ responds to $E_x E_z$, and $\sigma_{yx}$ responds to $E_x^3$. 

\begin{table}[htb]
\centering
\renewcommand{\arraystretch}{1.5}
\setlength{\tabcolsep}{0.0mm}{
\caption{\label{tab:compensation}{ The effective magnetic fields associated with several MPGs. On this condition, the application of bi-axial or tri-axial electric field along specific crystallographic directions might result in quenched effective magnetic fields. The quenching conditions are provided in the last column.}}
 \begin{tabular}{m{1.5cm}<{\centering}|m{5.4cm}<{\centering}|m{1.7cm}<{\centering}}
    \toprule
MPGs&Effective magnetic fields&Quenching\\
\hline
$4mm.1$&$B^{z}_\mathrm{eff}:E_xE_y(E_x^2-E_y^2)$&$\frac{|E_x|}{|E_y|}=1$\\
\hline
$\bar{4}2m.1$&$B^{z,xy}_\mathrm{eff}:E_xE_y(E_x^2-E_y^2)$&$\frac{|E_x|}{|E_y|}=1$\\
\hline
$4/mmm.1$&$B^{z,xy}_\mathrm{eff}:E_xE_y(E_x^2-E_y^2)$&$\frac{|E_x|}{|E_y|}=1$\\
\hline
$4/m^\prime mm$&$B^{z,xyz}_\mathrm{eff}:E_xE_yE_z(E_x^2-E_y^2)$&$\frac{|E_x|}{|E_y|}=1$\\
\hline
\multirow{2}*{$\bar{3}m.1$}&\multirow{2}*{$B^{z,xy}_\mathrm{eff}:E_xE_y(E_x^4-\frac{10}{3}E_x^2E_y^2+E_y^4)$}&$\frac{|E_x|}{|E_y|}=\frac{1}{\sqrt{3}}$ \\
~&~& $\frac{|E_y|}{|E_x|}=\frac{1}{\sqrt{3}}$\\
\hline
\multirow{2}*{$6mm.1$}&\multirow{2}*{$B^{z,xy}_\mathrm{eff}:E_xE_y(E_x^4-\frac{10}{3}E_x^2E_y^2+E_y^4)$}&$\frac{|E_x|}{|E_y|}=\frac{1}{\sqrt{3}}$\\
~&~& $\frac{|E_y|}{|E_x|}=\frac{1}{\sqrt{3}}$\\
\hline
\multirow{2}*{$6/mmm.1$}&\multirow{2}*{$B^{z,xy}_\mathrm{eff}:E_xE_y(E_x^4-\frac{10}{3}E_x^2E_y^2+E_y^4)$}&$\frac{|E_x|}{|E_y|}=\frac{1}{\sqrt{3}}$\\
~&~& $\frac{|E_y|}{|E_x|}=\frac{1}{\sqrt{3}}$\\
\hline
\multirow{2}*{$6/m^\prime m m$}&\multirow{2}*{$B^{z,xyz}_\mathrm{eff}:E_xE_yE_z(E_x^4-\frac{10}{3}E_x^2E_y^2+E_y^4)$}&$\frac{|E_x|}{|E_y|}=\frac{1}{\sqrt{3}}$\\
~&~& $\frac{|E_y|}{|E_x|}=\frac{1}{\sqrt{3}}$\\
\hline
\multirow{3}*{$\bar{4}3m.1$}&$B^{x,yz}_\mathrm{eff}:E_y^3E_z-E_yE_z^3$&$\frac{|E_y|}{|E_z|}=1$\\
\cline{2-3}
~&$B^{y,xz}_\mathrm{eff}:E_x^3E_z-E_xE_z^3$&$\frac{|E_x|}{|E_z|}=1$\\
\cline{2-3}
~&$B^{z,xy}_\mathrm{eff}:E_x^3E_y-E_xE_y^3$&$\frac{|E_x|}{|E_y|}=1$\\
\hline
\multirow{3}*{$m\bar{3}m.1$}&$B^{x,yz}_\mathrm{eff}:E_y^3E_z-E_yE_z^3$&$\frac{|E_y|}{|E_z|}=1$\\
\cline{2-3}
~&$B^{y,xz}_\mathrm{eff}:E_x^3E_z-E_xE_z^3$&$\frac{|E_x|}{|E_z|}=1$\\
\cline{2-3}
~&$B^{z,xy}_\mathrm{eff}:E_x^3E_y-E_xE_y^3$&$\frac{|E_x|}{|E_y|}=1$\\
\hline
\hline
\end{tabular}}
\end{table}

Before finishing this section, it is important to discuss several special cases regarding Table~\ref{tab:coupling1} of the Appendix A. An example is given by the $4mm.1$ MPG, for which $B^{z,xy}_\mathrm{eff}=\lambda_{zxy,31}E_xE_y(E_x^2 - E_y^2)$. As a matter of fact, the $\sigma_{yx}$ conductivity is induced by bi-axial $[E_x,E_y]$ electric field, via fourth-order magnetoelectric coupling $E_x^3 E_y$ or $E_x E_y^3$. Note, however, that $E_x^3 E_y$ and $E_x E_y^3$ have equal rights toward $\sigma_{yx}$. When $\frac{|E_x|}{|E_y|}=1$, the effective $B^{z,xy}_\mathrm{eff}$ becomes zero and this yields null $\sigma_{yx}$ conductivity ($\sigma_{yx}$ being quenched) in the regime of fourth-order magnetoelectric coupling. Such a quenching behavior appears in 10 MPGs on the condition that bi-axial or tri-axial electric field is along specific directions, as summarized in Table~\ref{tab:compensation}.\\

\subsection{Electric-field driven AHE in Cr$_2$O$_3$ and CoF$_2$}

We now explore semiconductors that exhibit electric-field driven AHE. As shown in Table~\ref{tab:SymmetryBreaking}, a vast majority of MPGs enable such a feature and materials belonging to these MPGs can be identified thanks to the MAGNDATA database~\cite{gallego2016magndata,gallego2016magndata2}. From the database, we select Cr$_2$O$_3$ and CoF$_2$ as two representative materials. These two materials have no more than 10 ions per primitive cell, for which the computational costs are affordable.

\begin{figure}[!h]
\includegraphics[width=1.\linewidth]{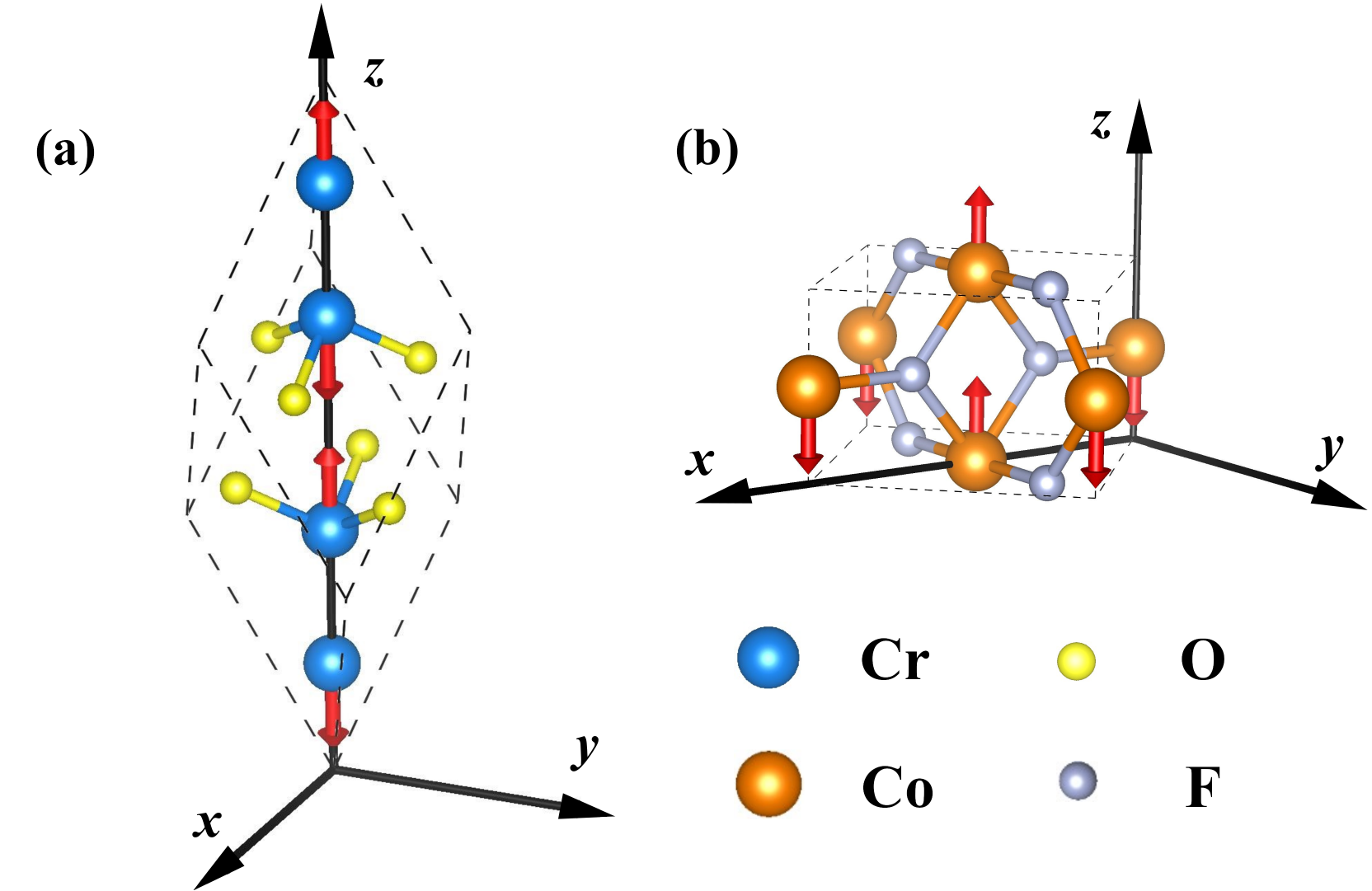}
\caption{\label{fig:stru} The crystal and magnetic structures for two compounds. The Cr, Co, O and F ions are denoted by cyan, orange, yellow and grey spheres, respectively. The magnetic moments are shown by red arrows.  (a): Cr$_2$O$_3$ with $+L$ magnetic order parameter. (b): CoF$_2$ with $+L$ magnetic order parameter. The $x$, $y$, and $z$ directions in each panel are orthogonal to each other.}
\end{figure}

\begin{figure*}[htb]
\includegraphics[width=1.\linewidth]{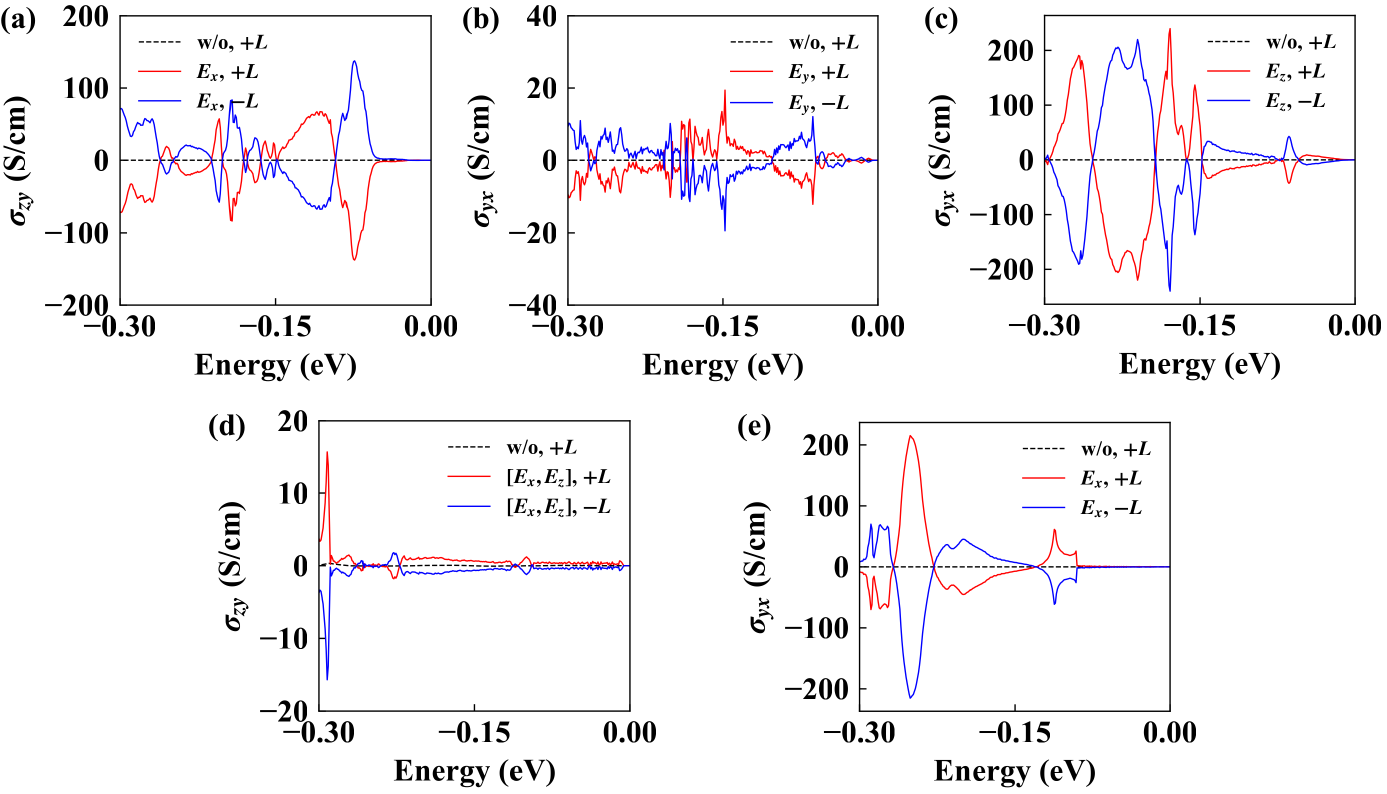}
\caption{\label{fig:AHE} The electric-field driven anomalous Hall conductivity components in Cr$_2$O$_3$ and CoF$_2$ as functions of chemical potential (defined with respect to the Fermi level).  Panels (a)-(c) depict the anomalous Hall conductivity components in Cr$_2$O$_3$: $\sigma_{zy}$ induced by a uni-axial $E_x$ field, $\sigma_{yx}$ induced by a uni-axial $E_y$ field, and $\sigma_{yx}$ induced by a uni-axial $E_z$ field. Panels (d) and (e) show the anomalous Hall conductivity components in CoF$_2$: $\sigma_{zy}$ induced by a bi-axial $[E_x,E_z]$ field, and $\sigma_{yx}$ induced by a uni-axial $E_x$ field.}
\end{figure*}

Cr$_2$O$_3$ has a N\'{e}el temperature of 307~K and belongs to the $\bar{3}^\prime m^\prime$ MPG~\cite{brown2002determination,mahmood2021voltage}. The crystal and magnetic structures of Cr$_2$O$_3$ are sketched in Fig.~\ref{fig:stru}(a). According to Table~\ref{tab:coupling1} of the Appendix A, Cr$_2$O$_3$ with $+L$ magnetic order parameter enables (i) $\sigma_{zy}$ conductivity driven by $E_x$, (ii) $\sigma_{xz}$ conductivity driven by $E_y$, and (iii) $\sigma_{yx}$ conductivity driven by $E_z$ or $E_y^3$. Furthermore, switching the magnetic order parameter from $+L$ to $-L$ reverses the aforementioned electric-field induced Hall conductivity, namely, $\sigma_{\alpha\beta}(+L) = -\sigma_{\alpha\beta}(-L)$ [see Eq.~(\ref{onsager})]. This is verified by our first-principles calculations of various anomalous Hall conductivity components for Cr$_2$O$_3$, as shown in Figs.~\ref{fig:AHE}(a)-(c) and Fig. S1(a) of the SM. To be specific, $\sigma_{zy}$, $\sigma_{xz}$, and $\sigma_{yx}$ components remain zero in the absence of an external electric field; Such components are driven by $E_x$, $E_y$, or $E_z$ electric field, where the induced conductivities associated with $+L$ and $-L$ are basically opposite by sign. As shown in Figs.~\ref{fig:AHE}(a),~\ref{fig:AHE}(c) and Fig. S1(a) of the SM, $\sigma_{zy}$, $\sigma_{xz}$, and $\sigma_{yx}$ can be larger than $100$ S/cm. These conductivities are driven by $E_x$, $E_y$, or $E_z$ electric field (magnitude being $4$ MV/cm) via first-order response. Instead, the maximum value of $\sigma_{yx}$, driven by $E_y = 4$~MV/cm via third-order response, is less than $20$ S/cm [see Fig.~\ref{fig:AHE}(b)]. Interestingly, Fig.~\ref{fig:AHE}(a) and Fig.~S1(a) of the SM are quite similar, and this is understandable by examining the effective magnetic fields associated with $\bar{3}^\prime m^\prime$ MPG. In Table~\ref{tab:coupling1} of the Appendix A, we find that $B^{x,x}_\mathrm{eff}=\lambda_{xx,1}E_x$ and $B^{y,y}_\mathrm{eff}=\lambda_{yy,1}E_y$ for $\bar{3}^\prime m^\prime$ MPG, where the relation $\lambda_{xx,1}=\lambda_{yy,1}$ implies that $\sigma_{zy}$ and $\sigma_{xz}$ are symmetrically related.

Another interesting material is CoF$_2$ which has a N\'{e}el temperature of 39 K~\cite{thomson2014cof2,metzger2024magnon,mashkovich2021terahertz}. In Fig.~\ref{fig:stru}(b), we schematically show the crystal and magnetic structures for CoF$_2$. Under such a coordinate system, the MPG of CoF$_2$ is $4^\prime/mm^\prime m$ --- being the MPG for the $P4^\prime/mmm^\prime$ magnetic space group~\cite{jauch2004gamma,PhysRevB.109.224312}. As shown in Table~\ref{tab:coupling1} of the Appendix A, unlike Cr$_2$O$_3$, the anomalous Hall conductivities in CoF$_2$ respond to electric field via second order. Specifically,  $\sigma_{yx}$ is driven by uni-axial $E_x^2$ or $E_y^2$ response, while $\sigma_{zy}$ and $\sigma_{xz}$ are from bi-axial $E_xE_z$ and $E_yE_z$ responses, respectively. Our first-principles calculations illustrated in Fig.~\ref{fig:AHE}(d), Fig.~\ref{fig:AHE}(e), Fig.~S1(b) of the SM, and Fig.~S1(c) of the SM, corroborate our aforementioned analysis and further confirm that reversing the magnetic order parameter (in CoF$_2$) reverses the anomalous Hall conductivities. For $E_x=E_z=2\sqrt{2}$~MV/cm and $E_y=E_z=2\sqrt{2}$~MV/cm (i.e., the total amplitude being $4$~MV/cm), $\sigma_{zy}$ and $\sigma_{xz}$ can reach 10~S/cm, whereas an electric field of $E_x=4$~MV/cm or $E_y=4$~MV/cm can yield $\sigma_{yx}$ being larger than 200~S/cm. Moreover, the resemblance between Figs.~\ref{fig:AHE}(d), (e) and Figs.~S1(b),~S1(c) of the SM can be attributed to the effective magnetic fields permitted by the $4^\prime/mm^\prime m$ MPG (see Table~\ref{tab:coupling1}). Associated with Fig.~\ref{fig:AHE}(d) and Fig.~S1(b) of the SM, the effective fields are $B^{x,xz}_\mathrm{eff}=\lambda_{xxz,11}E_xE_z$ and $B^{y,yz}_\mathrm{eff}=\lambda_{yyz,11}E_yE_z$ with $\lambda_{xxz,11}=-\lambda_{yyz,11}$. As for the other pair of figures, the effective magnetic fields are $B^{z,x}_\mathrm{eff}=\lambda_{zx,2}E_x^2$ and $B^{z,y}_\mathrm{eff}=\lambda_{zy,2}E_y^2$ with $\lambda_{zx,2}=-\lambda_{zy,2}$.\\

\subsection{Possible measurement strategies}

Finally, we provide two strategies for experimentally measuring the aforementioned electric-field induced AHE. Our proposed phenomena involve two types of electric fields, namely, a relatively large electric field $\mathbf{E}$ to polarize the material, and a tiny electric field $\boldsymbol{\mathcal{E}}$ to drive the Hall current density $\mathbf{J}$. When $\mathbf{E}$ is perpendicular to the plane defined by $\boldsymbol{\mathcal{E}}$ and $\mathbf{J}$, the experimental detection of the $\mathbf{E}$-driven AHE is straightforward. 
For instance, we can apply an out-of-plane gate field $\mathbf{E}$ and an in-plane bias field $\boldsymbol{\mathcal{E}}$ to induce the Hall current. This is exemplified by the experimental measurement of layer Hall effect in MnBi$_2$Te$_4$ (see e.g., Ref.~\cite{gao2021layer}). If $\mathbf{E}$ lies within the plane defined by $\boldsymbol{\mathcal{E}}$ and $\mathbf{J}$, the measurement becomes complicated because both $\mathbf{E}$ and $\boldsymbol{\mathcal{E}}$ might create anomalous Hall response. In such a case, we may use the ``$dc+ac$'' method (see e.g., Ref.~\cite{ye2025current}) to measure the $\mathbf{E}$-driven AHE. To be specific, we use a static field $\mathbf{E}_\mathrm{dc}$ to polarize the material and a alternating field $\boldsymbol{\mathcal{E}}_\mathrm{ac}$ to drive the Hall response~\cite{ye2025current}. This setup allows us to distinguish the anomalous Hall current induced by $\boldsymbol{\mathcal{E}}_\mathrm{ac}$ from background current associated with $\mathbf{E}_\mathrm{dc}$. The resulting alternating Hall current density $\mathbf{J}_\mathrm{ac}$, oscillating at the same frequency as $\boldsymbol{\mathcal{E}}_\mathrm{ac}$, can then be measured.

\section{Summary and outlook}

In the present work, we establish the phenomenological theory for the anomalous Hall effect in magnets. By symmetry analysis, we show that the anomalous Hall conductivities in magnets can either spontaneously appear or be driven by electric fields via linear or nonlinear magnetoelectric coupling. Such anomalous Hall conductivities are governed by magnetic point group symmetries and are hosted by 90 type-I and type-III magnetic point groups (summarized in Tables~\ref{tab:SymmetryBreaking} and~\ref{tab:coupling1}). To testify our theory, we further propose Cr$_2$O$_3$ and CoF$_2$ (by first-principles calculations) as candidate materials showcasing electric-field induced anomalous Hall conductivities. Strikingly, the $+L$ and $-L$ magnetic order parameters in magnets yield spontaneous or electric-field induced anomalous Hall conductivities with opposite signs. As a matter of fact, the probe of magnetic orders is of practical importance for spintronics~\cite{shao2020nonlinear,wang2023intrinsic,jungwirth2018multiple,baltz2018antiferromagnetic,nvemec2018antiferromagnetic,han2023coherent,wadley2016electrical,vzelezny2018spin,PhysRevLett.133.096803,saidl2017optical}. Magnetic orders can be detected by, for instance, anisotropic magnetoresistance~\cite{wadley2016electrical,vzelezny2018spin}, nonreciprocal charge transport~\cite{zhao2024general,zhao2025electrically,chen2022role}, magneto-optical Kerr effect~\cite{nvemec2018antiferromagnetic,saidl2017optical}, and nonlinear Hall effect~\cite{liu2021intrinsic,wang2023intrinsic,wang2021intrinsic,shao2020nonlinear}. Compared with these approaches, our work offers alternative avenues for the detection of magnetic orders in magnets.\\

\begin{acknowledgements}
The authors acknowledge the support from National Natural Science Foundation of China (Grants No. 12274174, No. 12274102, No. 22090044, No. 52288102, No. 52090024, and No. 12034009), the Strategic
373 Priority Research Program of Chinese Academy of Sciences (Grant No. XDB33000000). L.B. acknowledges support from the Vannevar Bush Faculty Fellowship (VBFF) from the Department of Defense and Award No. DMR-1906383 from the National Science Foundation Q-AMASE-i Program (MonArk NSF Quantum Foundry). L.J.Y. thanks the support from the high-performance computing center of Jilin University. H.J.Z. thanks the ``Xiaomi Young Scholar'' Proiect.\\
\end{acknowledgements}

\appendix
\section{Derivations of effective magnetic fields}
The effective $B^{\alpha,0}_\mathrm{eff}$, $B^{\alpha,\beta}_\mathrm{eff}$, $B^{\alpha,\beta\gamma}_\mathrm{eff}$, and $B^{\alpha,\beta\gamma\delta}_\mathrm{eff}$ fields are derived with respect to MPGs. In the following, we use $m^\prime m2^\prime$ and $4.1$ MPGs to demonstrate our derivations. First of all, we derive the energetic couplings between $M_\alpha$ and $E_\beta$ ($\alpha,\beta=x,y,z$) associated with $m^\prime m2^\prime$ MPG. These couplings should be invariant under all symmetry operations associated with the $m^\prime m2^\prime$ MPG, as required by symmetry. As shown in Table~\ref{tab:SymmetryBreaking}, $m^\prime m2^\prime$ MPG enables (i) a spontaneous $M_y$ component, (ii) an $M_z$ component driven by uni-axial $E_y$ electric field, and (iii) an $M_x$ component driven by bi-axial $[E_x,E_y]$ electric field. Following this, we heuristically guess that the energetic couplings (for $m^\prime m2^\prime$ MPG) might contain $\kappa_{y}M_y$, $\kappa_{zy,l} M_z E_y^l$, and $\kappa_{xxy,lm} M_x E_x^l E_y^m$ terms, with $\kappa_{y}$, $\kappa_{zy,l}$ and $\kappa_{xxy,lm}$ being coupling coefficients. The couplings $\kappa_{y}M_y$, $\kappa_{zy,1} M_z E_y$, and $\kappa_{xxy,11} M_x E_x E_y$ are invariant with respect to the symmetry operations of $m^\prime m2^\prime$ MPG. Similarly, $\kappa_{zy,3} M_z E_y^3$, $\kappa_{zy,5} M_z E_y^5$, $\kappa_{xxy,31} M_x E_x^3 E_y$, and $\kappa_{xxy,13} M_x E_x E_y^3$ and some other terms are invariant as well; Yet, these terms are higher-order responses to electric field and are thus neglected. Hence, for $m^\prime m2^\prime$ MPG, the $[E_x,E_y]$-driven $M_x$, spontaneous $M_y$, and $E_y$-driven $M_z$ correspond to
\begin{align}\label{energymm2}
&\Delta E^{x,xy}(m^\prime m2^\prime) = \kappa_{xxy,11} M_x E_x E_y,\notag\\
&\Delta E^{y,0}(m^\prime m2^\prime) = \kappa_{y}M_y,\\
&\Delta E^{z,y}(m^\prime m2^\prime) = \kappa_{zy,1} M_z E_y.\notag
\end{align}
Eq.~(\ref{energymm2}) suggests the effective magnetic fields as
\begin{align}\label{fieldmm2}
B^{x,xy}_\mathrm{eff}&(m^\prime m2^\prime)\notag\\ 
&\propto \frac{\partial{\Delta E^{x,xy}(m^\prime m2^\prime)}}{\partial{M_x}}|_{E_x\ne0,E_y\ne0,E_z=0}\notag\\ 
&= \kappa_{xxy,11} E_x E_y,\notag\\
B^{y,0}_\mathrm{eff}&(m^\prime m2^\prime)\notag\\
&\propto \frac{\partial{\Delta E^{y,0}(m^\prime m2^\prime)}}{\partial{M_y}}|_{E_x=0,E_y=0,E_z=0}\\  
&= \kappa_{y},\notag\\
B^{z,y}_\mathrm{eff}&(m^\prime m2^\prime)\notag\\
&\propto\frac{\partial{\Delta E^{z,y}(m^\prime m2^\prime)}}{\partial{M_z}}|_{E_x=0,E_y\ne0,E_z=0}\notag\\ 
&= \kappa_{zy,1} E_y.\notag
\end{align}

We move on to derive the effective magnetic field for the $4.1$ MPG. As shown in Table~\ref{tab:SymmetryBreaking}, the $4.1$ MPG enables (i) a spontaneous $M_z$ component, (ii) an $M_x$ component driven by uni-axial $E_x$ or $E_y$ electric field, and (iii) an $M_y$ component driven by uni-axial $E_x$ or $E_y$ electric field. The energetic couplings for $4.1$ MPG are thus guessed as $\kappa_{z}M_z$, $\kappa_{xx,n}M_xE_x^n$, $\kappa_{xy,n}M_xE_y^n$, $\kappa_{yx,n}M_yE_x^n$ and $\kappa_{yy,n}M_yE_y^n$ terms, with $n$ being positive integers. Working with $n=1$, we find that $\kappa_{z}M_z$ is invariant, while $\kappa_{xx,1}M_xE_x$, $\kappa_{yy,1}M_yE_y$, $\kappa_{xy,1}M_xE_y$, and $\kappa_{yx,1}M_yE_x$ are not for the $4.1$ MPG. For instance, $\mathfrak{4}^{+}_z$ and $\mathfrak{4}^{-}_z$ transform $\kappa_{xx,1}M_xE_x$ to $\kappa_{xx,1}M_yE_y$ and $\kappa_{yy,1}M_y E_y$ to $\kappa_{yy,1}M_x E_x$. As a matter of fact, $\mathfrak{4}^{+}_z$ and $\mathfrak{4}^{-}_z$ symmetry operations link $M_x E_x$ and $M_y E_y$, and the corresponding invariant couplings are $\kappa_{xx,1}(M_xE_x+M_yE_y)$ and $\kappa_{yy,1}(M_yE_y+M_xE_x)$ with $\kappa_{xx,1}=\kappa_{yy,1}$. Similarly, we find that $\kappa_{xy,1}(M_xE_y-M_yE_x)$ and $\kappa_{yx,1}(M_yE_x-M_xE_y)$ with $\kappa_{xy,1}=-\kappa_{yx,1}$ are also invariant couplings. Neglecting higher order terms (e.g., $n>1$), the energetic couplings for $4.1$ MPG are
\begin{align}\label{energy41}
&\Delta E^{x,x}(4.1) = \kappa_{xx,1}(M_xE_x+M_yE_y),\notag\\
&\Delta E^{x,y}(4.1) = \kappa_{xy,1}(M_xE_y-M_yE_x),\notag\\
&\Delta E^{y,y}(4.1) = \kappa_{yy,1}(M_yE_y+M_xE_x),\\
&\Delta E^{y,x}(4.1) = \kappa_{yx,1}(M_yE_x-M_xE_y),\notag\\
&\Delta E^{z,0}(4.1) = \kappa_{z}M_z,\notag
\end{align}
which yields the effective fields as
\begin{align}\label{field41}
B^{x,x}_\mathrm{eff}(4.1)&\propto \frac{\partial{\Delta E^{x,x}(4.1)}}{\partial{M_x}}|_{E_x\ne0,E_y=0,E_z=0}\notag\\
&=\kappa_{xx,1}E_x \notag\\
B^{x,y}_\mathrm{eff}(4.1)&\propto \frac{\partial{\Delta E^{x,y}(4.1)}}{\partial{M_x}}|_{E_x=0,E_y\ne0,E_z=0}\notag\\
&=\kappa_{xy,1}E_y,\notag\\
B^{y,y}_\mathrm{eff}(4.1)&\propto \frac{\partial{\Delta E^{y,y}(4.1)}}{\partial{M_y}}|_{E_x=0,E_y\ne0,E_z=0}\\
&= \kappa_{yy,1}E_y,\notag\\
B^{y,x}_\mathrm{eff}(4.1)&\propto\frac{\partial{\Delta E^{y,x}(4.1)}}{\partial{M_y}}|_{E_x\ne0,E_y=0,E_z=0}\notag\\
&=\kappa_{yx,1}E_x,\notag\\
B^{z,0}_\mathrm{eff}(4.1)&\propto \frac{\partial{\Delta E^{z,0}(4.1)}}{\partial{M_z}}|_{E_x=0,E_y=0,E_z=0}\notag\\
&= \kappa_{z}.\notag
\end{align}
According to Eq.~(\ref{h_alpha_expand}), we rewrite Eqs.~(\ref{fieldmm2}) and~(\ref{field41}) as
\begin{align}\label{beffmm241}
&B^{x,xy}_\mathrm{eff}(m^\prime m2^\prime) = \lambda_{xxy,11} E_x E_y,\notag\\
&B^{y,0}_\mathrm{eff}(m^\prime m2^\prime) = \lambda_{y},\notag\\
&B^{z,y}_\mathrm{eff}(m^\prime m2^\prime) = \lambda_{zy,1} E_y,\notag \\
&B^{x,x}_\mathrm{eff}(4.1) = \lambda_{xx,1}E_x,\\
&B^{x,y}_\mathrm{eff}(4.1) = \lambda_{xy,1}E_y,\notag\\
&B^{y,y}_\mathrm{eff}(4.1) = \lambda_{yy,1}E_y,\notag\\
&B^{y,x}_\mathrm{eff}(4.1) = \lambda_{yx,1}E_x,\notag\\
&B^{z,0}_\mathrm{eff}(4.1)= \lambda_{z}.\notag
\end{align}
In Eq.~(\ref{beffmm241}), the effective fields of $4.1$ MPG satisfy the relations $\lambda_{xx,1}=\lambda_{yy,1}$ and $\lambda_{xy,1}=-\lambda_{yx,1}$ due to the $\mathfrak{4}^{+}_z$ and $\mathfrak{4}^{-}_z$ symmetry operations. By similar procedures, we derive the effective magnetic fields for the other type-I and type-III MPGs, summarized in Table~\ref{tab:coupling1}.

\twocolumngrid
\begin{widetext}
{\renewcommand{\arraystretch}{1.5}
\tablecaption{The effective magnetic fields associated with type-I and type-III MPGs. For an MPG, the $B^{\alpha,0}_\mathrm{eff}$, $B^{\alpha,\beta}_\mathrm{eff}$, $B^{\alpha,\beta\gamma}_\mathrm{eff}$ or $B^{\alpha,\beta\gamma\delta}_\mathrm{eff}$ effective magnetic field is listed in the corresponding $B^{\alpha}_\mathrm{eff}$ entry. In some entries, different effective magnetic fields are separated by semicolons. Additionally, some effective magnetic fields can be related by symmetry, which are shown in the last column.\label{tab:coupling1}}
\tablefirsthead{
\hline \hline
MPG&$B^{x}_\mathrm{eff}$&$B^{y}_\mathrm{eff}$&$B^{z}_\mathrm{eff}$&Relationship\\
\hline
}
\tablehead{
\multicolumn{5}{c}{\tablename\ \thetable{} -- Continued from previous page}\\
\hline \hline
MPG&$B^{x}_\mathrm{eff}$&$B^{y}_\mathrm{eff}$&$B^{z}_\mathrm{eff}$&Relationship\\
\hline
}
\tabletail{
\hline
\multicolumn{5}{r}{{Continued on next page}} \\
\hline
}
\tablelasttail{
\hline
\hline
}
{
\footnotesize
\begin{supertabular}{m{1.4cm}<{\centering}m{3.8cm}<{\centering}m{3.8cm}<{\centering}m{3.8cm}<{\centering}m{4.5cm}<{\centering}}
$1.1$&$\lambda_x$&$\lambda_y$&$\lambda_z$&$/$\\
\hline
$\bar{1}.1$&$\lambda_x$&$\lambda_y$&$\lambda_z$&$/$\\
\hline
$\bar{1}^\prime$&$\lambda_{xx,1}E_x;\lambda_{xy,1}E_y;\lambda_{xz,1}E_z$&$\lambda_{yx,1}E_x;\lambda_{yy,1}E_y;\lambda_{yz,1}E_z$&$\lambda_{zx,1}E_x;\lambda_{zy,1}E_y;\lambda_{zz,1}E_z$&$/$\\
\hline
$2.1$&$\lambda_{xx,1}E_x;\lambda_{xy,1}E_y$&$\lambda_{yx,1}E_x;\lambda_{yy,1}E_y$&$\lambda_z$&$/$\\
\hline
$2^\prime$&$\lambda_x$&$\lambda_y$&$\lambda_{zx,1}E_x;\lambda_{zy,1}E_y$&$/$\\
\hline
$m.1$&$\lambda_{xz,1}E_z$&$\lambda_{yz,1}E_z$&$\lambda_z$&$/$\\
\hline
$m^\prime$&$\lambda_x$&$\lambda_y$&$\lambda_{zz,1}E_z$&$/$\\
\hline
$2/m.1$&$\lambda_{xxz,11}E_xE_z;\lambda_{xyz,11}E_yE_z$&$\lambda_{yxz,11}E_xE_z;\lambda_{yyz,11}E_yE_z$&$\lambda_z$&$/$\\
\hline
$2^\prime/m$&$\lambda_{xz,1}E_z$&$\lambda_{yz,1}E_z$&$\lambda_{zx,1}E_x;\lambda_{zy,1}E_y$&$/$\\
\hline
$2/m^\prime$&$\lambda_{xx,1}E_x;\lambda_{xy,1}E_y$&$\lambda_{yx,1}E_x;\lambda_{yy,1}E_y$&$\lambda_{zz,1}E_z$&$/$\\
\hline
$2^\prime/m^\prime$&$\lambda_x$&$\lambda_y$&$\lambda_{zxz,11}E_xE_z;\lambda_{zyz,11}E_yE_z$&$/$\\
\hline
$222.1$&$\lambda_{xx,1}E_x$&$\lambda_{yy,1}E_y$&$\lambda_{zz,1}E_z$&$/$\\
\hline
$2^\prime2^\prime 2$&$\lambda_{xy,1}E_y$&$\lambda_{yx,1}E_x$&$\lambda_z$&$/$\\
\hline
$mm2.1$&$\lambda_{xy,1}E_y$&$\lambda_{yx,1}E_x$&$\lambda_{zxy,11}E_xE_y$&$/$\\
\hline
$m^\prime m2^\prime$&$\lambda_{xxy,11}E_xE_y$&$\lambda_y$&$\lambda_{zy,1}E_y$&$/$\\
\hline
$m^\prime m^\prime2$&$\lambda_{xx,1}E_x$&$\lambda_{yy,1}E_y$&$\lambda_z$&$/$\\
\hline
$mmm.1$&$\lambda_{xyz,11}E_yE_z$&$\lambda_{yxz,11}E_xE_z$&$\lambda_{zxy,11}E_xE_y$&$/$\\
\hline
$m^\prime mm$&$\lambda_{xxyz,111}E_xE_yE_z$&$\lambda_{yz,1}E_z$&$\lambda_{zy,1}E_y$&$/$\\
\hline
$m^\prime m^\prime m$&$\lambda_{xxz,11}E_xE_z$&$\lambda_{yyz,11}E_yE_z$&$\lambda_z$&$/$\\
\hline
$m^\prime m^\prime m^\prime$&$\lambda_{xx,1}E_x$&$\lambda_{yy,1}E_y$&$\lambda_{zz,1}E_z$&$/$\\
\hline
$4.1$&$\lambda_{xx,1}E_x;\lambda_{xy,1}E_y$&$\lambda_{yy,1}E_y;\lambda_{yx,1}E_x$&$\lambda_z$&$\lambda_{xx,1}=\lambda_{yy,1};\lambda_{xy,1}=-\lambda_{yx,1}$\\
\hline
\multirow{2}*{$4^\prime$}&\multirow{2}*{$\lambda_{xx,1}E_x;\lambda_{xy,1}E_y$}&\multirow{2}*{$\lambda_{yy,1}E_y;\lambda_{yx,1}E_x$}&\multirow{2}*{$\lambda_{zx,2}E_x^2;\lambda_{zy,2}E_y^2$}&\multirow{2}*{\tabincell{m{4.5cm}<{\centering}}{$\lambda_{xx,1}=-\lambda_{yy,1};\lambda_{xy,1}=\lambda_{yx,1};$\\$\lambda_{zx,2}=-\lambda_{zy,2}$}}\\
~&~&~&~&~\\
\hline
$\bar{4}.1$&$\lambda_{xx,1}E_x;\lambda_{xy,1}E_y$&$\lambda_{yy,1}E_y;\lambda_{yx,1}E_x$&$\lambda_z$&$\lambda_{xx,1}=-\lambda_{yy,1};\lambda_{xy,1}=\lambda_{yx,1}$\\
\hline
\multirow{2}*{$\bar{4}^\prime$}&\multirow{2}*{$\lambda_{xx,1}E_x;\lambda_{xy,1}E_y$}&\multirow{2}*{$\lambda_{yy,1}E_y;\lambda_{yx,1}E_x$}&\multirow{2}*{$\lambda_{zx,2}E_x^2;\lambda_{zy,2}E_y^2;\lambda_{zz,1}E_z$}&\multirow{2}*{\tabincell{m{4.5cm}<{\centering}}{$\lambda_{xx,1}=\lambda_{yy,1};\lambda_{xy,1}=-\lambda_{yx,1}$\\$\lambda_{zx,2}=-\lambda_{zy,2}$}}\\
~&~&~&~&~\\
\hline
\multirow{2}*{$4/m.1$}&\multirow{2}*{$\lambda_{xxz,11}E_xE_z;\lambda_{xyz,11}E_yE_z$}&\multirow{2}*{$\lambda_{yxz,11}E_xE_z;\lambda_{yyz,11}E_yE_z$}&\multirow{2}*{$\lambda_z$}&\multirow{2}*{\tabincell{m{4.5cm}<{\centering}}{$\lambda_{xxz,11}=\lambda_{yyz,11};$\\$\lambda_{xyz,11}=-\lambda_{yxz,11}$}}\\
~&~&~&~&~\\
\hline
\multirow{2}*{$4^\prime/m$}&\multirow{2}*{$\lambda_{xxz,11}E_xE_z;\lambda_{xyz,11}E_yE_z$}&\multirow{2}*{$\lambda_{yxz,11}E_xE_z;\lambda_{yyz,11}E_yE_z$}&\multirow{2}*{$\lambda_{zx,2}E_x^2;\lambda_{zy,2}E_y^2$}&\multirow{2}*{\tabincell{m{4.5cm}<{\centering}}{$\lambda_{xxz,11}=-\lambda_{yyz,11};$\\$\lambda_{xyz,11}=\lambda_{yxz,11};\lambda_{zx,2}=-\lambda_{zy,2}$}}\\
~&~&~&~&~\\
\hline
$4/m^\prime$&$\lambda_{xx,1}E_x;\lambda_{xy,1}E_y$&$\lambda_{yx,1}E_x;\lambda_{yy,1}E_y$&$\lambda_{zz,1}E_z$&$\lambda_{xx,1}=\lambda_{yy,1};\lambda_{xy,1}=-\lambda_{yx,1}$\\
\hline
\multirow{2}*{$4^\prime/m^\prime$}&\multirow{2}*{$\lambda_{xx,1}E_x;\lambda_{xy,1}E_y$}&\multirow{2}*{$\lambda_{yx,1}E_x;\lambda_{yy,1}E_y$}&\multirow{2}*{$\lambda_{zxz,21}E_x^2E_z;\lambda_{zyz,21}E_y^2E_z$}&\multirow{2}*{\tabincell{m{4.5cm}<{\centering}}{$\lambda_{xx,1}=-\lambda_{yy,1};\lambda_{xy,1}=\lambda_{yx,1}$\\$\lambda_{zxz,21}=-\lambda_{zyz,21}$}}\\
~&~&~&~&~\\
\hline
$422.1$&$\lambda_{xx,1}E_x$&$\lambda_{yy,1}E_y$&$\lambda_{zz,1}E_z$&$\lambda_{xx,1}=\lambda_{yy,1}$\\
\hline
\multirow{2}*{$4^\prime 2 2^\prime$}&\multirow{2}*{$\lambda_{xx,1}E_x$}&\multirow{2}*{$\lambda_{yy,1}E_y$}&\multirow{2}*{\tabincell{m{3.8cm}<{\centering}}{$\lambda_{zxy,11}E_xE_y;\lambda_{zxz,21}E_x^2E_z;$\\$\lambda_{zyz,21}E_y^2E_z$}}&\multirow{2}*{\tabincell{m{4.5cm}<{\centering}}{$\lambda_{xx,1}=-\lambda_{yy,1};$\\$\lambda_{zxz,21}=-\lambda_{zyz,21}$}}\\
~&~&~&~\\
\hline
$4 2^\prime 2^\prime$&$\lambda_{xy,1}E_y$&$\lambda_{yx,1}E_x$&$\lambda_z$&$\lambda_{xy,1}=-\lambda_{yx,1}$\\
\hline
$4mm.1$&$\lambda_{xy,1}E_y$&$\lambda_{yx,1}E_x$&$\lambda_{zxy,31}E_xE_y(E_x^2-E_y^2)$&$\lambda_{xy,1}=-\lambda_{yx,1}$\\
\hline
$4^\prime m^\prime m$&$\lambda_{xx,1}E_x$&$\lambda_{yy,1}E_y$&$\lambda_{zx,2}E_x^2;\lambda_{zy,2}E_y^2$&$\lambda_{xx,1}=-\lambda_{yy,1};\lambda_{zx,2}=-\lambda_{zy,2}$\\
\hline
$4 m^\prime m^\prime$&$\lambda_{xx,1}E_x$&$\lambda_{yy,1}E_y$&$\lambda_z$&$\lambda_{xx,1}=\lambda_{yy,1}$\\
\hline
\multirow{2}*{$\bar{4}2m.1$}&\multirow{2}*{$\lambda_{xx,1}E_x$}&\multirow{2}*{$\lambda_{yy,1}E_y$}&\multirow{2}*{\tabincell{m{3.8cm}<{\centering}}{$\lambda_{zxy,31}E_xE_y(E_x^2-E_y^2);$\\$\lambda_{zxz,21}E_x^2E_z;\lambda_{zyz,21}E_y^2E_z$}}&\multirow{2}*{\tabincell{m{4.5cm}<{\centering}}{$\lambda_{xx,1}=-\lambda_{yy,1};$\\$\lambda_{zxz,21}=-\lambda_{zyz,21}$}}\\
~&~&~&~\\
\hline
$\bar{4}^\prime2^\prime m$&$\lambda_{xy,1}E_y$&$\lambda_{yx,1}E_x$&$\lambda_{zx,2}E_x^2;\lambda_{zy,2}E_y^2$&$\lambda_{yx,1}=-\lambda_{xy,1};\lambda_{zx,2}=-\lambda_{zy,2}$\\
\hline
$\bar{4}^\prime2m^\prime$&$\lambda_{xx,1}E_x$&$\lambda_{yy,1}E_y$&$\lambda_{zz,1}E_z$&$\lambda_{xx,1}=\lambda_{yy,1}$\\
\hline
$\bar{4}2^\prime m^\prime$&$\lambda_{xy,1}E_y$&$\lambda_{yx,1}E_x$&$\lambda_z$&$\lambda_{xx,1}=\lambda_{yy,1}$\\
\hline
$4/mmm.1$&$\lambda_{xyz,11}E_yE_z$&$\lambda_{yxz,11}E_xE_z$&$\lambda_{zxy,31}E_xE_y(E_x^2-E_y^2)$&$\lambda_{xyz,11}=-\lambda_{yxz,11}$\\
\hline
$4/m^\prime mm$&$\lambda_{xy,1}E_y$&$\lambda_{yx,1}E_x$&$\lambda_{zxyz,311}E_xE_yE_z(E_x^2-E_y^2)$&$\lambda_{xy,1}=-\lambda_{yx,1}$\\
\hline
\multirow{2}*{$4^\prime/mm^\prime m$}&\multirow{2}*{$\lambda_{xxz,11}E_xE_z$}&\multirow{2}*{$\lambda_{yyz,11}E_yE_z$}&\multirow{2}*{$\lambda_{zx,2}E_x^2;\lambda_{zy,2}E_y^2$}&\multirow{2}*{\tabincell{m{4.5cm}<{\centering}}{$\lambda_{xxz,11}=\lambda_{yyz,11};$\\$\lambda_{zx,2}=-\lambda_{zy,2}$}}\\
~&~&~&~&~\\
\hline
\multirow{2}*{$4^\prime/m^\prime m^\prime m$}&\multirow{2}*{$\lambda_{xx,1}E_x$}&\multirow{2}*{$\lambda_{yy,1}E_y$}&\multirow{2}*{$\lambda_{zxz,21}E_x^2E_z;\lambda_{zyz,21}E_y^2E_z$}&\multirow{2}*{\tabincell{m{4.5cm}<{\centering}}{$\lambda_{xx,1}=-\lambda_{yy,1};$\\$\lambda_{zxz,21}=-\lambda_{zyz,21}$}}\\
~&~&~&~&~\\
\hline
$4/mm^\prime m^\prime$&$\lambda_{xxz,11}E_xE_z$&$\lambda_{yyz,11}E_yE_z$&$\lambda_z$&$\lambda_{xxz,11}=\lambda_{yyz,11}$\\
\hline
$4/m^\prime m^\prime m^\prime$&$\lambda_{xx,1}E_x$&$\lambda_{yy,1}E_y$&$\lambda_{zz,1}E_z$&$\lambda_{xx,1}=\lambda_{yy,1}$\\
\hline
$3.1$&$\lambda_{xx,1}E_x;\lambda_{xy,1}E_y$&$\lambda_{yy,1}E_y;\lambda_{yx,1}E_x$&$\lambda_z$&$\lambda_{xx,1}=\lambda_{yy,1};\lambda_{xy,1}=-\lambda_{yx,1}$\\
\hline
$\bar{3}.1$&$\lambda_{xx,2}E_x^2;\lambda_{xy,2}E_y^2$&$\lambda_{yx,2}E_x^2;\lambda_{yy,2}E_y^2$&$\lambda_z$&$\lambda_{xx,2}=-\lambda_{xy,2};\lambda_{yx,2}=-\lambda_{yy,2}$\\
\hline
$\bar{3}^\prime$&$\lambda_{xx,1}E_x;\lambda_{xy,1}E_y$&$\lambda_{yy,1}E_y;\lambda_{yx,1}E_x$&$\lambda_{zx,3}E_x^3;\lambda_{zy,3}E_y^3;\lambda_{zz}E_z$&$\lambda_{xx,1}=\lambda_{yy,1};\lambda_{xy,1}=-\lambda_{yx,1}$\\
\hline
$32.1$&$\lambda_{xx,1}E_x;\lambda_{xy,2}E_y^2$&$\lambda_{yy,1}E_y$&$\lambda_{zy,3}E_y^3;\lambda_{zz,1}E_z$&$\lambda_{xx,1}=\lambda_{yy,1}$\\
\hline
$32^\prime$&$\lambda_{xy,1}E_y$&$\lambda_{yx,1}E_x;\lambda_{yy,2}E_y^2$&$\lambda_z$&$\lambda_{xy,1}=-\lambda_{yx,1}$\\
\hline
$3m.1$&$\lambda_{xy,1}E_y;\lambda_{xx,2}E_x^2$&$\lambda_{yx,1}E_x$&$\lambda_{zx,3}E_x^3$&$\lambda_{xy,1}=-\lambda_{yx,1}$\\
\hline
$3m^\prime$&$\lambda_{xx,1}E_x$&$\lambda_{yy,1}E_y;\lambda_{yx,2}E_x^2$&$\lambda_z$&$\lambda_{xx,1}=-\lambda_{yy,1}$\\
\hline
\multirow{2}*{$\bar{3}m.1$}&\multirow{2}*{$\lambda_{xx,2}E_x^2;\lambda_{xy,2}E_y^2$}&\multirow{2}*{\tabincell{m{3.cm}<{\centering}}{$\lambda_{yxy,11}E_xE_y;$\\$\lambda_{yxz,11}E_xE_z$}}&\multirow{2}*{\tabincell{m{4.cm}<{\centering}}{$\lambda_{zxy,33}E_xE_y(3E_x^4-10E_x^2$\\$E_y^2+3E_y^4);\lambda_{zxz,31}E_x^3E_z$}}&\multirow{2}*{$\lambda_{xx,2}=-\lambda_{xy,2}=-\frac{1}{2}\lambda_{yxy,11}$}\\
~&~&~&~\\
\hline
$\bar{3}^\prime m$&$\lambda_{xy,1}E_y$&$\lambda_{yx,1}E_x$&$\lambda_{zx,3}E_x^3$&$\lambda_{xy,1}=-\lambda_{yx,1}$\\
\hline
$\bar{3}^\prime m^\prime$&$\lambda_{xx,1}E_x$&$\lambda_{yy,1}E_y$&$\lambda_{zy,3}E_y^3;\lambda_{zz,1}E_z$&$\lambda_{xx,1}=\lambda_{yy,1}$\\
\hline
\multirow{2}*{$\bar{3}m^\prime$}&\multirow{2}*{\tabincell{m{3.cm}<{\centering}}{$\lambda_{xxy,11}E_xE_y;$\\$\lambda_{xxz,11}E_xE_z$}}&\multirow{2}*{$\lambda_{yx,2}E_x^2;\lambda_{yy,2}E_y^2$}&\multirow{2}*{$\lambda_z$}&\multirow{2}*{$\lambda_{yx,2}=-\lambda_{yy,2}=\frac{1}{2}\lambda_{xxy,11}$}\\
~&~&~&~\\
\hline
$6.1$&$\lambda_{xx,1}E_x;\lambda_{xy,1}E_y$&$\lambda_{yy,1}E_y;\lambda_{yx,1}E_x$&$\lambda_z$&$\lambda_{xx,1}=\lambda_{yy,1};\lambda_{xy,1}=-\lambda_{yx,1}$\\
\hline
\multirow{2}*{$6^\prime$}&\multirow{2}*{$\lambda_{xx,2}E_x^2;\lambda_{xy,2}E_y^2$}&\multirow{2}*{$\lambda_{yy,2}E_y^2;\lambda_{yx,2}E_x^2$}&\multirow{2}*{$\lambda_{zx,3}E_x^3;\lambda_{zy,3}E_y^3$}&\multirow{2}*{\tabincell{m{4.5cm}<{\centering}}{$\lambda_{xx,2}=-\lambda_{xy,2};$\\$\lambda_{yy,2}=-\lambda_{yx,2}$}}\\
~&~&~&~\\
\hline
\multirow{2}*{$\bar{6}.1$}&\multirow{2}*{\tabincell{m{3.cm}<{\centering}}{$\lambda_{xxz,11}E_xE_z;$\\$\lambda_{xyz,11}E_yE_z$}}&\multirow{2}*{\tabincell{m{3.cm}<{\centering}}{$\lambda_{yyz,11}E_yE_z;$\\$\lambda_{yxz,11}E_xE_z$}}&\multirow{2}*{$\lambda_z$}&\multirow{2}*{\tabincell{m{4.5cm}<{\centering}}{$\lambda_{xxz,11}=\lambda_{yyz,11}$\\$\lambda_{yxz,11}=-\lambda_{xyz,11}$}}\\
~&~&~&~\\
\hline
$\bar{6}^\prime$&$\lambda_{xx,1}E_x;\lambda_{xy,1}E_y$&$\lambda_{yy,1}E_y;\lambda_{yx,1}E_x$&$\lambda_{zz,1}E_z$&$\lambda_{xx,1}=\lambda_{yy,1};\lambda_{xy,1}=-\lambda_{yx,1}$\\
\hline
\multirow{2}*{$6/m.1$}&\multirow{2}*{\tabincell{m{3.cm}<{\centering}}{$\lambda_{xxz,11}E_xE_z;$\\$\lambda_{xyz,11}E_yE_z$}}&\multirow{2}*{\tabincell{m{3.cm}<{\centering}}{$\lambda_{yyz,11}E_yE_z;$\\$\lambda_{yxz,11}E_xE_z$}}&\multirow{2}*{$\lambda_z$}&\multirow{2}*{\tabincell{m{4.5cm}<{\centering}}{$\lambda_{xxz,11}=\lambda_{yyz,11};$\\$\lambda_{xyz,11}=-\lambda_{yxz,11}$}}\\
~&~&~&~\\
\hline
\multirow{2}*{$6^\prime/m$}&\multirow{2}*{\tabincell{m{3.cm}<{\centering}}{$\lambda_{xxz,21}E_x^2E_z;$\\$\lambda_{xyz,21}E_y^2E_z$}}&\multirow{2}*{\tabincell{m{3.cm}<{\centering}}{$\lambda_{yyz,21}E_y^2E_z;$\\$\lambda_{yxz,21}E_x^2E_z$}}&\multirow{2}*{$\lambda_{zx,3}E_x^3;\lambda_{zy,3}E_y^3$}&\multirow{2}*{\tabincell{m{3.cm}<{\centering}}{$\lambda_{xxz,21}=-\lambda_{xyz,21};$\\$\lambda_{yyz,21}=-\lambda_{yxz,21}$}}\\
~&~&~&~\\
\hline
$6/m^\prime$&$\lambda_{xx,1}E_x;\lambda_{xy,1}E_y$&$\lambda_{yy,1}E_y;\lambda_{yx,1}E_x$&$\lambda_{zz,1}E_z$&$\lambda_{xx,1}=\lambda_{yy,1};\lambda_{xy,1}=-\lambda_{yx,1}$\\
\hline
$6^\prime/m^\prime$&$\lambda_{xx,2}E_x^2;\lambda_{xy,2}E_y^2$&$\lambda_{yy,2}E_y^2;\lambda_{yx,2}E_x^2$&$\lambda_{zxz,31}E_x^3E_z;\lambda_{zyz,31}E_y^3E_z$&$\lambda_{xx,2}=-\lambda_{xy,2};\lambda_{yy,2}=-\lambda_{yx,2}$\\
\hline
$622.1$&$\lambda_{xx,1}E_x$&$\lambda_{yy,1}E_y$&$\lambda_{zz,1}E_z$&$\lambda_{xx,2}=\lambda_{yy,2}$\\
\hline
\multirow{3}*{$6^\prime22^\prime$}&\multirow{3}*{$\lambda_{xx,2}E_x^2;\lambda_{xy,2}E_y^2$}&\multirow{3}*{\tabincell{m{3.cm}<{\centering}}{$\lambda_{yxy,11}E_xE_y;$\\$\lambda_{yxz,21}E_x^2E_z;$\\$\lambda_{yyz,21}E_y^2E_z$}}&\multirow{3}*{$\lambda_{zy,3}E_y^3$}&\multirow{3}*{\tabincell{m{4.5cm}<{\centering}}{$\lambda_{xx,2}=-\lambda_{xy,2}=\frac{1}{2}\lambda_{yxy,11};$\\$\lambda_{yxz,21}=-\lambda_{yyz,21}$}}\\
~&~&~&~\\
~&~&~&~\\
\hline
$62^\prime 2^\prime$&$\lambda_{xy,1}E_y$&$\lambda_{yx,1}E_x$&$\lambda_z$&$\lambda_{xy,1}=-\lambda_{yx,1}$\\
\hline
\multirow{2}*{$6mm.1$}&\multirow{2}*{$\lambda_{xy,1}E_y$}&\multirow{2}*{$\lambda_{yx,1}E_x$}&\multirow{2}*{\tabincell{m{3.cm}<{\centering}}{$\lambda_{zxy,33}E_xE_y(3E_x^4-$\\$10E_x^2E_y^2+3E_y^4)$}}&\multirow{2}*{$\lambda_{xy,1}=-\lambda_{yx,1}$}\\
~&~&~&~\\
\hline
$6^\prime m m^\prime$&$\lambda_{xx,2}E_x^2;\lambda_{xy,2}E_y^2$&$\lambda_{yxy,11}E_xE_y$&$\lambda_{zx,3}E_x^3$&$\lambda_{xx,2}=-\lambda_{xy,2}=-\frac{1}{2}\lambda_{yxy,11}$\\
\hline
$6m^\prime m^\prime$&$\lambda_{xx,1}E_x$&$\lambda_{yy,1}E_y$&$\lambda_z$&$\lambda_{xx,1}=\lambda_{yy,1}$\\
\hline
\multirow{2}*{$\bar{6}m2.1$}&\multirow{2}*{\tabincell{m{3.cm}<{\centering}}{$\lambda_{xxz,21}E_x^2E_z;$\\$\lambda_{xyz,11}E_yE_z$}}&\multirow{2}*{$\lambda_{yxz,11}E_xE_z$}&\multirow{2}*{$\lambda_{zx,3}E_x^3$}&\multirow{2}*{$\lambda_{yxz,11}=-\lambda_{xyz,11}$}\\
~&~&~&~\\
\hline
$\bar{6}^\prime m^\prime 2$&$\lambda_{xx,1}E_x$&$\lambda_{yy,1}E_y;\lambda_{yx,2}E_x^2$&$\lambda_{zz,1}E_z$&$\lambda_{xx,1}=\lambda_{yy,1}$\\
\hline
$\bar{6}^\prime m2^\prime$&$\lambda_{xx,2}E_x^2;\lambda_{xy,1}E_y$&$\lambda_{yx,1}E_x$&$\lambda_{zxz,31}E_x^3E_z$&$\lambda_{yx,1}=-\lambda_{xy,1}$\\
\hline
\multirow{2}*{$\bar{6}m^\prime2^\prime$}&\multirow{2}*{$\lambda_{xxz,11}E_xE_z$}&\multirow{2}*{\tabincell{m{3.cm}<{\centering}}{$\lambda_{yyz,11}E_yE_z;$\\$\lambda_{yxz,21}E_x^2E_z$}}&\multirow{2}*{$\lambda_z$}&\multirow{2}*{$\lambda_{xxz,11}E_xE_z=\lambda_{yyz,11}$}\\
~&~&~&~\\
\hline
\multirow{2}*{$6/mmm.1$}&\multirow{2}*{$\lambda_{xyz,11}E_yE_z$}&\multirow{2}*{$\lambda_{yxz,11}E_xE_z$}&\multirow{2}*{\tabincell{m{3.cm}<{\centering}}{$\lambda_{zxy,33}E_xE_y(3E_x^4-$\\$10E_x^2E_y^2+3E_y^4)$}}&\multirow{2}*{$\lambda_{xyz,11}=-\lambda_{yxz,11}$}\\
~&~&~&~\\
\hline
\multirow{2}*{$6/m^\prime m m$}&\multirow{2}*{$\lambda_{xy,1}E_y$}&\multirow{2}*{$\lambda_{yx,1}E_x$}&\multirow{2}*{\tabincell{m{3.cm}<{\centering}}{$\lambda_{zxyz,331}E_xE_yE_z(3E_x^4-$\\$10E_x^2E_y^2+3E_y^4)$}}&\multirow{2}*{$\lambda_{xy,1}=-\lambda_{yx,1}$}\\
~&~&~&~\\
\hline
\multirow{2}*{$6^\prime/mmm^\prime$}&\multirow{2}*{\tabincell{m{3.cm}<{\centering}}{$\lambda_{xxz,21}E_x^2E_z;$\\$\lambda_{xyz,21}E_y^2E_z$}}&\multirow{2}*{$\lambda_{yxyz,111}E_xE_yE_z$}&\multirow{2}*{$\lambda_{zx,3}E_x^3$}&\multirow{2}*{\tabincell{m{3.cm}<{\centering}}{$\lambda_{xxz,21}=-\lambda_{xyz,21}$}}\\
~&~&~&~\\
\hline
$6^\prime/m^\prime mm^\prime$&$\lambda_{xx,2}E_x^2;\lambda_{xy,2}E_y^2$&$\lambda_{yxy,11}E_xE_y$&$\lambda_{zxz,31}E_x^3E_z$&$\lambda_{xx,2}=-\lambda_{xy,2}=-\frac{1}{2}\lambda_{yxy,11}$\\
\hline
$6/mm^\prime m^\prime$&$\lambda_{xxz,11}E_xE_z$&$\lambda_{yyz,11}E_yE_z$&$\lambda_z$&$\lambda_{xxz,11}=\lambda_{yyz,11}$\\
\hline
$6/m^\prime m^\prime m^\prime$&$\lambda_{xx,1}E_x$&$\lambda_{yy,1}E_y$&$\lambda_{zz,1}E_z$&$\lambda_{xx,1}=\lambda_{yy,1}$\\
\hline
$23.1$&$\lambda_{xx,1}E_x$&$\lambda_{yy,1}E_y$&$\lambda_{zz,1}E_z$&$\lambda_{xx,1}=\lambda_{yy,1}=\lambda_{zz,1}$\\
\hline
$432.1$&$\lambda_{xx,1}E_x$&$\lambda_{yy,1}E_y$&$\lambda_{zz,1}E_z$&$\lambda_{xx,1}=\lambda_{yy,1}=\lambda_{zz,1}$\\
\hline
\multirow{3}*{$4^\prime32^\prime$}&\multirow{3}*{\tabincell{m{3.8cm}<{\centering}}{$\lambda_{xxy,12}E_xE_y^2;\lambda_{xxz,12}E_xE_z^2;$\\$\lambda_{xyz,11}E_yE_z$}}&\multirow{3}*{\tabincell{m{3.8cm}<{\centering}}{$\lambda_{yyx,12}E_yE_x^2;\lambda_{yyz,21}E_yE_z^2;$\\$\lambda_{yxz,11}E_xE_z$}}&\multirow{3}*{\tabincell{m{3.8cm}<{\centering}}{$\lambda_{zzx,12}E_zE_x^2;\lambda_{zzy,12}E_zE_y^2;$\\$\lambda_{zxy,11}E_xE_y$}}&\multirow{3}*{\tabincell{m{4.5cm}<{\centering}}{$\lambda_{xxy,12}=-\lambda_{xxz,12}=-\lambda_{yyx,12}=$\\$\lambda_{yyz,21}=\lambda_{zzx,12}=-\lambda_{zxy,11};$\\$\lambda_{xyz,11}=\lambda_{yxz,11}=\lambda_{zxy,11}$}}\\
~&~&~&~\\
~&~&~&~\\
\hline
\multirow{3}*{$\bar{4}3m.1$}&\multirow{3}*{\tabincell{m{3.8cm}<{\centering}}{$\lambda_{xxy,12}E_xE_y^2;\lambda_{xxz,12}E_xE_z^2;$\\$\lambda_{xyz,31}(E_y^3E_z-E_yE_z^3)$}}&\multirow{3}*{\tabincell{m{3.8cm}<{\centering}}{$\lambda_{yyx,12}E_yE_x^2;\lambda_{yyz,12}E_yE_z^2;$\\$\lambda_{yzx,31}(E_z^3E_x-E_zE_x^3)$}}&\multirow{3}*{\tabincell{m{3.8cm}<{\centering}}{$\lambda_{zzx,12}E_zE_x^2;\lambda_{zzy,12}E_zE_y^2;$\\$\lambda_{zxy,31}(E_x^3E_y-E_xE_y^3)$}}&\multirow{3}*{\tabincell{m{4.5cm}<{\centering}}{$\lambda_{xxy,12}=-\lambda_{xxz,12}=-\lambda_{yyx,12}=$\\$\lambda_{yyz,21}=\lambda_{zzx,12}=-\lambda_{zxy,11};$\\$\lambda_{xyz,31}=\lambda_{yzx,31}=\lambda_{zxy,31}$}}\\
~&~&~&~\\
~&~&~&~\\
\hline
$\bar{4}^\prime3m^\prime$&$\lambda_{xx,1}E_x$&$\lambda_{yy,1}E_y$&$\lambda_{zz,1}E_z$&$\lambda_{xx,1}=\lambda_{yy,1}=\lambda_{zz,1}$\\
\hline
$m\bar{3}.1$&$\lambda_{xyz,11}E_yE_z$&$\lambda_{yxz,11}E_xE_z$&$\lambda_{zxy,11}E_xE_y$&$\lambda_{xyz,11}=\lambda_{yxz,11}=\lambda_{zxy,11}$\\
\hline
$m^\prime \bar{3}^\prime$&$\lambda_{xx,1}E_x$&$\lambda_{yy,1}E_y$&$\lambda_{zz,1}E_z$&$\lambda_{xx,1}=\lambda_{yy,1}=\lambda_{zz,1}$\\
\hline
$m\bar{3}m.1$&$\lambda_{xyz,31}(E_y^3E_z-E_yE_z^3)$&$\lambda_{yzx,31}(E_z^3E_x-E_zE_x^3)$&$\lambda_{zxy,31}(E_x^3E_y-E_xE_y^3)$&$\lambda_{xyz,31}=\lambda_{yzx,31}=\lambda_{zxy,31}$\\
\hline
\multirow{2}*{$m^\prime\bar{3}^\prime m$}&\multirow{2}*{\tabincell{m{3.8cm}<{\centering}}{$\lambda_{xxy,12}E_xE_y^2;\lambda_{xxz,12}E_xE_z^2;$}}&\multirow{2}*{\tabincell{m{3.8cm}<{\centering}}{$\lambda_{yyx,12}E_yE_x^2;\lambda_{yyz,12}E_yE_z^2;$}}&\multirow{2}*{\tabincell{m{3.8cm}<{\centering}}{$\lambda_{zzx,12}E_zE_x^2;\lambda_{zzy,12}E_zE_y^2;$}}&\multirow{2}*{\tabincell{m{4.5cm}<{\centering}}{$\lambda_{xxy,12}=-\lambda_{xxz,12}=-\lambda_{yyx,12}=$\\$\lambda_{yyz,12}=\lambda_{zzx,12}=-\lambda_{zxy,12};$}}\\
~&~&~&~\\
\hline
$m\bar{3}m^\prime$&$\lambda_{xyz,11}E_yE_z$&$\lambda_{yxz,11}E_xE_z$&$\lambda_{zxy,11}E_xE_y$&$\lambda_{xyz,11}=\lambda_{yxz,11}=\lambda_{zxy,11}$\\
\hline
$m^\prime\bar{3}^\prime m^\prime$&$\lambda_{xx,1}E_x$&$\lambda_{yy,1}E_y$&$\lambda_{zz,1}E_z$&$\lambda_{xx,1}=\lambda_{yy,1}=\lambda_{zz,1}$\\
\hline

\end{supertabular}
}
~\\
~\\
}
\end{widetext}


\input{main.bbl}

\end{document}

%% file: main.bbl
%